# FERMIONIC ZERO MODES FOR DYONS AND CHIRAL SYMMETRY BREAKING IN QCD

A.Gonzalez-Arroyo$^{(a)}$ and Yu.A.Simonov$^{(b)}$

(a) Depto. Física Teórica C-XI,
Universidad Autónoma de Madrid,
Cantoblanco, Madrid 28049 , SPAIN

(b) Institute of Theoretical and Experimental Physics,
Moscow 117259, Bolshaya Cheremushkinskaya, 25, Russia

## Abstract

Dyonic classical solutions of Yang-Mills theory are considered and the complete set of fermionic zero modes of these solutions are studied. Representing the QCD vacuum as a gas of dyons, one obtains chiral symmetry breaking due to zero modes similarly to the case of instantonic vacuum.

## 1   Introduction

Dyonic solutions (DS) are known already for a long time [1-4] and were mostly used in the context of the QCD vacuum at nonzero temperature [5]. Using a (singular) gauge transformation one can present DS in a form where its magnetic and electric fields are time-independent and the former displays a clear magnetic monopole character [6].

One can also visualize DS in the 't Hooft ansatz [7,8] as a coherent infinite chain of instantons placed equidistantlly along the time axis.

The case of a finite number of instantons was also considered [9] and it was shown that the magnetic monopole field is acting over a distance $R$ approximately equal to the length of instanton chain $L$.

The long distance field of DS can in principle bring about confinement, and simple estimates [9] show that a stochastic ensemble of dyons, e.g. noninteracting gas, can produce an area law of Wilson loop, in agreement with the idea of the dual Meissner effect [10].

Recently dyons appeared in $N = 2$ SUSY Yang–Mills model [11] and it was shown that the strong coupling region corresponds in the model to the weak coupling theory of dyons and monopoles. Condensation of those produces confinement again in line with the dual Meissner effect [10]. It is instructive to compare properties of DS and those of instantons. It is known that in the dilute gas of instantons confinement does not occur [12], nevertheless instantons may play significant role in the QCD vacuum.

One important property of instantons is that it gives rise to fermionic zero modes (FZM) [13] and this brings in a lot of interesting physical consequences, e.g. the famous 't Hooft's determinantal interaction [14] yields proton instability, instantonic solution of the U(1) problem [15], and finally chiral symmetry breaking (CSB) in the instanton gas or liquid model [16-17]. It is the latter property which was studied analytically in [18-21] and recent numerical analysis of instantonic model vacuum [22] suggests that many realistic properties of $q\bar{q}$ correlators in the model are due to a reasonable description of CSB.

A natural question arises whether DS can also ensure CSB. To this end we study in this paper fermionic zero modes of DS. For some cases some FZM are known explicitly, e.g. the Jackiw-Rebbi zero modes [23]. In our case an infinite set of FZM was found long ago [24].

We present in this paper a concise and economic formalism for DS and explicit form of its FZM. We also find a correspondence between one of the FZM and its Jackiw-Rebbi form. This we do in the first 3 chapters.

In the rest of the paper we exploit the FZM of DS to study the question of CSB. To this end we consider a gas of dyons and apply the technique of instanton gas model to derive the effective propagation for quarks in the gas of dyons. We specifically study the formation of chiral (constituent) quark mass and chiral condensate in the dyonic gas and compare results with those of instanton gas model.

The plan of the paper is as follows. In the next Section we adopt useful notations and economic formalism to rewrite gauge field in the 't Hooft ansatz and the corresponding Dirac equation for the quark zero modes. In Section 3 we define exact dyonic solutions and study its properties. In Section 4 we classify fermionic zero modes on a dyon, and find correspondence with the Jackiw-Rebbi solutions [23]. In Section 5 we define the gas of dyons and in Section 6 we modify the formalism of the instanton gas model [18,21] to study CSB in the gas of dyons and antidyons. The last section is devoted to the discussion of results and perspectives of dyonic gas model.

## 2  Notation and formalism

We begin our consideration by setting the notation. We are working in the Euclidean space–time, with indices running from 0 to 3. Summing over repeated indices is implied. Let us define matrices $\Gamma_\mu$ and $\bar{\Gamma}_\mu$ as follows:

$$\Gamma_\mu \equiv (I, \ -i\vec{\sigma}) \tag{1}$$

$$\bar{\Gamma}_\mu \equiv (I, \ i\vec{\sigma})$$

where $\vec{\sigma}$ are the Pauli matrices. The $\eta$ and $\bar{\eta}$ symbols are then given by

$$\Gamma_\mu \bar{\Gamma}_\nu = \eta_{\alpha\mu\nu} \bar{\Gamma}_\alpha \tag{2}$$

$$\bar{\Gamma}_\mu \Gamma_\nu = \bar{\eta}_{\alpha\mu\nu} \Gamma_\alpha$$

where all indices take 4 values (0,1,2,3) and $\eta_{0\mu\nu} = \bar{\eta}_{0\mu\nu} = \delta_{\mu\nu}$ (our notation differs from that of 't Hooft). Given a real 4-vector $a_\mu$ we may define the matrices

$$\hat{a} = a_\mu \Gamma_\mu \tag{3}$$

$$\hat{\bar{a}} = a_\mu \bar{\Gamma}_\mu = \hat{a}^+$$

and indeed we have

$$\hat{a}\hat{\bar{a}} = a^2 \ I \tag{4}$$

with $a^2 = a_\mu a_\mu$. In what follows we will be considering both colour and spin indices and for that purpose we will write $\hat{a}_c$ or $\hat{a}_s$ depending on which space the matrices are acting on.

Fermionic fields $\psi_{\alpha a}(x)$ carry both colour ($a$) and spin indices ($\alpha$). The representation of the Clifford algebra that we are considering is

$$\gamma_\mu = \begin{pmatrix} 0 & \Gamma_\mu \\ \bar{\Gamma}_\mu & 0 \end{pmatrix}, \quad \gamma_5 = \begin{pmatrix} 1 & 0 \\ 0 & -1 \end{pmatrix}. \tag{5}$$

In what follows we will concentrate on chiral fields $\gamma_5 \psi_\pm = \pm\psi_\pm$. The Dirac equation and operator for positive chirality fields $\psi_+$ reduces to

$$\hat{\bar{D}}\psi_+ = 0 \ , \quad \hat{\bar{D}} \equiv (\partial_\mu - iA_\mu)\bar{\Gamma}_{\mu s} \tag{6}$$

and for those of negative chirality $\psi_-$ is

$$\hat{D}\psi_- = 0 \ ; \quad \hat{D} = (\partial_\mu - iA_\mu)\Gamma_{\mu s} \ , \tag{7}$$

where $A_\mu$ is the vector potential matrix and subscript $s$ in $\Gamma_{\mu s}$ implies that SU(2) indices in $\Gamma$ refer to spin, whereas those in $\Gamma_{\mu c}$ – to colour.

In what follows we will concentrate ourselves on vector potentials of 't Hooft ansatz form [7,8]

$$A_\mu^a = -\bar{\eta}_{a\mu\nu}\partial_\nu f \ . \tag{8}$$

Defining $A_\mu^0 = \partial_\mu f$ we may introduce the matrix in color space

$$\mathcal{A}_\mu = \frac{i}{2} A_\mu^\alpha \Gamma_\alpha \tag{9}$$

which differs from $A_\mu$ written before by a term proportional to the unit matrix. In this form 't Hooft ansatz can be written as

$$\mathcal{A}_\mu = \bar{\Gamma}_{\mu c}(\hat{\partial}_c f)(-\frac{i}{2}) \tag{10}$$

where all matrices are acting on color space. In terms of this matrix $\mathcal{A}_\mu$ we may introduce modified Dirac operators

$$\hat{\mathcal{D}} \equiv (\partial_\mu - i\mathcal{A}_\mu)\Gamma_{\mu s} = \hat{D} - \frac{1}{2}(\hat{\partial}_s f) = e^{\frac{1}{2}f}\hat{D}e^{-\frac{1}{2}f} \tag{11}$$

$$\hat{\bar{\mathcal{D}}} \equiv (\partial_\mu - i\mathcal{A}_\mu)\bar{\Gamma}_{\mu s} = \hat{\bar{D}} - \frac{1}{2}(\hat{\bar{\partial}}_s f) = e^{\frac{1}{2}f}\hat{\bar{D}}e^{-\frac{1}{2}f}$$

An important role in what follows in played by the matrix $M$

$$M = \frac{1}{2}\sum_\mu \Gamma_{\mu s}\bar{\Gamma}_{\mu c} = \frac{1}{2}\sum_\mu \bar{\Gamma}_{\mu s}\Gamma_{\mu c} \tag{12}$$

more explicitly $M_{\alpha a;\beta b} \equiv \frac{1}{2}\sum_\mu (\Gamma_{\mu s})_{\alpha\beta}(\bar{\Gamma}_{\mu c})_{ab}$; which acts on the $2 \times 2$ dimensional space of chiral spinors. This matrix is self adjoint, unitary and satisfies the following relations.

$$M^2 = 1$$
$$\hat{a}_s M = M\hat{a}_c$$
$$\hat{a}_c M = M\hat{a}_s \tag{13}$$
$$\hat{\bar{a}}_s M = M\hat{\bar{a}}_c$$
$$\hat{\bar{a}}_c M = M\hat{\bar{a}}_s$$

From $M$ we can construct the projectors

$$P_+ = \frac{1}{2}(1 - M) = \frac{1}{4}\sum_\mu \Gamma_{\mu s}\Gamma_{\mu c} = \frac{1}{4}\sum_\mu \bar{\Gamma}_{\mu s}\bar{\Gamma}_{\mu c} \qquad (14)$$

$$P_- = \frac{1}{2}(1 + M)$$

which project onto a 1 and 3 dimensional space respectively. These spaces correspond to the total spin zero and total spin one spaces of the product of the original spin-isospin one half spaces. The dimensionality follows from $TrM = 2$. In the $V_+$ space, projected by $P_+$, all vectors are proportional to the unique element of the basis, $u_+$ which is taken to be unitary. With this notation the modified Dirac operators for 't Hooft ansatz are given by

$$\hat{\mathcal{D}} = \hat{\partial}_s - M(\hat{\partial}_c f) \qquad (15)$$

$$\hat{\bar{\mathcal{D}}} = \hat{\bar{\partial}}_s - 2P_+(\hat{\partial}_c f) \qquad (16)$$

We close this section by giving some relations which will useful later in finding solutions to the Dirac equations. We have

$$\Gamma_{\mu s}\, u_+ = \bar{\Gamma}_{\mu c}\, u_+ \qquad (17)$$

$$\Gamma_{\mu c}\, u_+ = \bar{\Gamma}_{\mu s}\, u_+$$

which can be proven by realizing that $(u_+)_{a\alpha} \propto (\sigma_2)_{a\alpha}$.

## 3 Euclidean dyons and antidyons.

Using the 't Hooft ansatz [7,8] formula one can compute the field strength tensor $F_{\mu\nu}$. Separating it into self-dual and anti-self-dual parts

$$F_{\mu\nu} = \eta_{i\mu\nu}\omega_i^{(+)} + \bar{\eta}_{i\mu\nu}\omega_i^{(-)} \qquad (18)$$

we get

$$\omega_i^{(-)} = -\frac{i}{4}\Gamma_{ic} W \Delta_4 W^{-1} \qquad (19)$$

$$\omega_i^{(+)} = -\frac{i}{4} W^{-1} \hat{\bar{\partial}} \bar{\Gamma}_{ic} \hat{\partial} W$$

where
$$F \equiv \frac{1}{W} \equiv e^{-f} \tag{20}$$

For self-dual solutions $\omega_i^{(-)}$ vanishes. As seen from Eq.(19), if $F_1$ and $F_2$ are two solutions of $\omega_i^{(-)} = 0$, then $F_1 + F_2$ is again a solution provided they do not have overlapping singularities. Hence one can obtain solutions by combining the pure-gauge solutions:

$$F = \lambda \tag{21}$$

$$F = \frac{\rho^2}{(x - x^0)^2}.$$

We have, thus the two main series of solutions:

$$F = 1 + \sum_{i=1}^{N} \frac{\rho_i^2}{(x - x^{(i)})^2} \tag{22}$$

$$F = \sum_{i=1}^{N} \frac{\rho_i^2}{(x - x^{(i)})^2} \tag{23}$$

Now we will specialize on dyonic solutions. These follow from the second of the above series (Eq. (23)) for $\rho_i = 1$ and $x^{(n)} = (\vec{0}, 2\pi n)$ ( we introduce dimensionless coordinates dividing $x_\mu$ and $x^{(n)}$ by some arbitrary scale parameter).

One obtains the form found by Harrington and Sheppard [2]

$$F_{HS} = \frac{1}{2r} \frac{\sinh r}{(\cosh r - \cos t)} \tag{24}$$

where $t \equiv x_0$ and $r \equiv |\vec{x}|$. The corresponding $A_{\mu a}$ is time dependent. There exists however a general singular gauge transformation [9], such that

$$\tilde{A}_\mu = U^+(A_\mu + i\partial_\mu)U; \quad U = exp(i\frac{\vec{\sigma}_c \vec{n}}{2}\theta), \tag{25}$$

where
$$tan\theta = W_0[\frac{W}{r} - W_r]^{-1}, \tag{26}$$

which in case of infinite multiinstanton chain (24) yields time–independent solutions (Rossi solutions [6])

$$\tilde{A}_{ia} = \frac{1}{r}\epsilon_{iab}n_b\left(\frac{r}{sinh\,r} - 1\right) \tag{27}$$

$$\tilde{A}_{0a} = \frac{1}{r}n_a(r\,coth\,r - 1) \tag{28}$$

The field $F_{\mu\nu}$ of the dyon is selfdual,

$$E_{ka} = B_{ka} = \delta_{ak}\frac{g'}{r} + n_a n_k\left(-\frac{g'}{r} + \frac{g^2 - 1}{r^2}\right) \tag{29}$$

where $g$ is

$$g = \frac{r}{sinh\,r} \tag{30}$$

For an anti-dyon solution, which is the gauge transform of the multi-antiinstanton solution, one has

$$\bar{A}_{ia} = \frac{1}{r}\epsilon_{iab}n_b\left(\frac{r}{sinh\,r} - 1\right) \tag{31}$$

$$\bar{A}_{0a} = -\frac{1}{r}n_a(r\,coth\,r - 1) \tag{32}$$

The magnetic monopole number of this solution $Q$ is opposite to that of the dyon.

The multidyon solutions are obtained from the 't Hooft ansatz (10)-(23) when one takes several parallel lines of instanton centers $x^{(i)}$. We shall not be using those explicitly below.

## 4  Fermionic zero modes on a dyon.

The set of fermionic zero modes is easily obtained from our formulas in section 2, specifically (6), (9), (11), (16). Here we present the result (see the Appendix for details of derivation). The set is one-parametrical, with real parameter $\beta : 0 < \beta \leq 1$

$$\psi^{(\beta)} = F^{1/2}\hat{\bar{\partial}}(F^{-1}F^{(\beta)})\,u_+ \tag{33}$$

where $F = F_{HS}$, given by (24), and

$$F^{(\beta)} = \sum_{n=-\infty}^{\infty} \frac{e^{i\beta n 2\pi}}{r^2 + (x_0 + 2\pi n)^2} \qquad (34)$$

This set of solutions can be connected to a set of zero modes found in [24]; for more details see the Appendix .

Summing up in (34) one obtains

$$F^{(\beta)} = \frac{e^{-it\beta}}{2r} \left( \frac{sinh((1-|\beta|)r) + cost \cdot sinh(r|\beta|) + isint\ sinh(r|\beta|)}{coshr - cost} \right) \qquad (35)$$

In particular for $\beta = \frac{1}{2}$ one has

$$F^{(\beta=\frac{1}{2})} = \frac{sinh\frac{r}{2} \cdot \cos\frac{t}{2}}{r(coshr - \cos t)} \qquad (36)$$

Insertion of (36) and (24) into (33) yields

$$\psi^{(\beta=\frac{1}{2})} = -\left( \frac{sinh\frac{r}{2}}{8r cosh^3 \frac{r}{2}} \right)^{\frac{1}{2}} U u_+ \qquad (37)$$

where $U$ is the inverse of the gauge transformation that yields the time-independent field strength (Eq. (25)), and given by

$$U = \sqrt{\frac{2}{coshr - \cos t}} \left( sin\frac{t}{2} cosh\frac{r}{2} + i(\vec{\sigma}_c \vec{n}) cos\frac{t}{2} sinh\frac{r}{2} \right) . \qquad (38)$$

Hence, the $\beta = \frac{1}{2}$ zero-mode can be gauged transformed to a time-independent solution. All other solutions (for $\beta \neq \frac{1}{2}$) are in general time dependent.

For the monopole classical solution it was shown by Jackiw and Rebbi, that there exists at least one normalizable (in 3d) zero fermionic mode $\chi$, which can be written as [23]

$$\chi = \frac{1}{r}(g(r) + h(r)(\vec{\sigma}_s \vec{n}))u_+ . \qquad (39)$$

The only normalizable solution has the form [23] $h \equiv 0$,

$$g(r) = const\ r\ exp[\int_0^r \frac{dr'}{r'}(A(r') - \frac{1}{2}B(r'))] \qquad (40)$$

where
$$A(r) = \frac{r}{sinh\, r} - 1 \; ; \;\; B(r) = r\, coth\, r - 1 \; . \tag{41}$$

Doing integrals in (39) one obtains
$$g(r) = const\sqrt{r}\frac{(1-e^{-r})^{1/2}}{(1+e^{-r})^{3/2}}e^{-\frac{r}{2}} = const\frac{\sqrt{r}(tanh\frac{r}{2})^{1/2}}{cosh\frac{r}{2}} \tag{42}$$

which is seen to give rise to the gauge transform of Eq. (37), demonstrating that the Jackiw-Rebbi zero mode coincides (up to a gauge transformation) with our solution $\psi^{(\beta=\frac{1}{2})}$.

## 5 Dyonic gas model of the QCD vacuum

We assume that the QCD vacuum at zero temperature can be realistically modelled by the dilute gas of dyons and antidyons. We start with an ansatz for gluonic fields:
$$A_\mu(x) = \sum_{i=1}^{N} A_\mu^{(i)}(x) + B_\mu(x) \tag{43}$$

where $B_\mu(x)$ comprises perturbative and other possible nondyonic fields with prescribed gauge transformation
$$B_\mu(x) \to U^+(x)(B_\mu(x) + i\, \partial_\mu)U(x) \tag{44}$$

Each of $A_\mu^{(i)}(x)$ is the field of dyon or antidyon which is made of solutions (27-28;31-32) by some shift $R^{(i)}$ and $0(4)$ rotation of the time direction $\omega^{(i)}$, and global color rotation $\Omega^{(i)}$
$$A_\mu^{(i)}(x) \equiv A_\mu^{(i)}(x, R^{(i)}, \omega^{(i)}, \Omega^{(i)}) \tag{45}$$

where for each unit 4 vector $\omega^{(i)}$ one can define
$$r = [(x-R)^2 - ((x_\mu - R_\mu)\omega_\mu)^2]^{1/2} \tag{46}$$
$$t = (x_\mu - R_\mu)\omega_\mu \tag{47}$$

so that (omitting superscripts $i$ everywhere)
$$A_\mu^{(i)}(x) = \Omega^{(i)+}\tilde{A}_\mu(r,t)\Omega^{(i)} \tag{48}$$

and $\tilde A_\mu$ is given by (8),(27),(28). One can see in (47) that the homogeneous gauge transformation for $A_\mu^{(i)}$ is ensured by the property of $\Omega$, that $\Omega \to \Omega U$, $\Omega^+ = U^+ \Omega^+$, and all construction in (43) is gauge covariant.

At this point one must specify in which gauge $A_\mu^{(i)}(x)$ in (43) are summed up. It is tempting to use the gauge ( we shall call it the Rossi gauge [6]), where $A_\mu^{(i)}$ is time-independent as in (27-28), representing static monopole solutions of the Prasad–Sommerfield type [3]. However, in this case the sum (43) leads to an infinite action, whenever all possible orientations $\omega^{(i)}$ can be present in the sum.

We have found only one gauge (modulo global rotations) where the action of the superposition (43) is finite. This is actually the singular gauge of the original 't Hooft ansatz (8),

$$A_{\mu a}^{(i)} = -\bar\eta_{a\mu\nu} \partial_\nu \ln W^{(i)},$$

$$(W^{(i)})^{-1} = \frac{1}{2r} \frac{sinh r}{cosh r - cos t} \qquad (49)$$

yielding for a standard (not shifted and not rotated) solution the form

$$A_{ia}^{(i)} = \epsilon_{aik} n_k \left(\frac{1}{r} - coth r + \frac{sinh r}{cosh r - cos t}\right) + \delta_{ia} \frac{sin t}{cosh r - cos t}$$

$$A_{0a}^{(i)} = -n_a \left(\frac{1}{r} - coth r + \frac{sinh r}{cosh r - cos t}\right) \qquad (50)$$

At large distances solutions (50) behave as

$$A_{\mu a}^{(i)} \sim \frac{1}{r}, \quad F_{\mu\nu}^{(i)} \sim \frac{1}{r^2} \qquad (51)$$

and the same is true for the sum (43).

Hence the total action of a gas of dyons of finite length is finite. To ensure that dyonic gas could be a realistic model of the QCD vacuum one must investigate the following points:

1) to check that the classical interaction between (anti) dyons is weak enough at large distances, so that the dilute gas approximation could be reasonably justified.

2) to prove the existence of the thermodynamic limit for the dyonic ensemble (42), i.e. that the total action of the (big) volume $V_4$ is proportional to the volume, when it increases.

One must also prove that the free energy of the $d\bar{d}$ calculated with quantum corrections has a minimum at a finite (and dilute) density.

All these points are now under investigation [25] and we shall not go into details, because of lack of space, but rather assume that the stable dilute $d\bar{d}$ gas indeed exists. The main question for us in the present paper is then whether the CSB occurs in such a gas. This point will be studied below in section 6.

# 6 Chiral symmetry breaking in the dyonic gas

For the gas made of equal number of dyons $N_+$ and antidyons $N_-$, $N_+ = N_- = \frac{N}{2}$ in the big volume $V_4$ we assume that the thermodynamic limit exists for the total action and other extensive quantities like the free energy, when $N \to \infty$, $V_4 \to \infty$ and $\frac{N}{V_4}$ is fixed and finite.

We shall use for the dyonic gas the formalism similar to that exploited for the instanton gas by Diakonov and Petrov [18].

The main problem to be solved in this section is: given fermionic zero modes on each of dyons and antidyons; find the full quark Green's function for the dyonic gas with the superposition ansatz (43).

To this end we make the same interpolating approximation for the one-dyon quark Green's function $S^{(i)}$ as in [18], i.e. in the exact spectral representation of $S^{(i)}$, $i = 1, \ldots N$

$$S^{(i)}(x, y) = \sum_n \frac{u_n^{(i)}(x) u_n^{(i)+}(y)}{\lambda_n - im} \qquad (52)$$

containing all modes $n = 1, 2, \ldots \infty$, we keep only zero modes $u_s^{(i)}$ and replace the nonzero-mode contribution by the free Green's function, since they coincide at large $n \sim \sqrt{p^2}$.

Thus with $S_0 = (-i\hat{D}(B) - im)^{-1}$ one has

$$S^{(i)}(x, y) = S_0(x, y) + \sum_{\text{zero modes}} \frac{u_s^{(i)}(x) u_s^{(i)+}(y)}{-im} \qquad (53)$$

One can see that $S^{(i)}$ diverges as $m \to 0$, but we shall show however that the total Green's function is finite for $m \to 0$ if $N_+ = N_-$.

Using (53) and (43) one derives the total Green's function as in [18,21] to be

$$S = S_0 - \sum_{\substack{i,k \\ n,m}} u_n^{(i)}(x) \left(\frac{1}{im + \hat{V}}\right)_{nm}^{ik} u_m^{(k)+}(y) \tag{54}$$

where upper indices $i, k$ run over all dyon numbers, $1 \leq i, k \leq N$, while lower indices $n, m$ run over all set of zero modes of the given dyon with the numbers $i, k$. We have also defined

$$V_{nm}^{ik} \equiv \int u_n^{(i)+}(x) i(\hat{\partial} - i\hat{B}) u_m^{(k)}(x) d^4x \tag{55}$$

We keep here the field $B$ to make the formalism gauge invariant; in estimates we systematically put $B_\mu$ equal to zero. Note that $u_n^+$ and $u_m$ in (55) should have opposite chiralities, hence $V^{ik}$ refer to dyon–antidyon $(d\bar{d})$ or opposite $(\bar{d}d)$ transitions, otherwise $V^{ik}$ is zero.

One can introduce graphs as in [18] to describe each term in (54) as a propagation amplitude from a dyon $i$ to a dyon $k$ through scattering on many intermediate (anti)dyons centers, with scattering amplitude of each center (dyon) being $\frac{1}{im}$ and transition amplitude from center $j$ (excited to the s-th level) to center $l$ (excited to the r-th level) being $V_{sr}^{jl}$.

The lower indices are not the only new element in (54-55) as compared to the instanton gas model [18-19]. The zero modes $u_n^{(i)}$ depend also on the Lorentz orientation $\omega^{(i)}$ of dyon, in addition to the color orientation $\Omega^{(i)}$ and position $R^{(i)}$ of the dyon, see Eq. (45).

$$u_n^{(i)}(x) \equiv \Omega^{(i)} u_n^{(i)}(x - R^{(i)}, \omega^{(i)}) \tag{56}$$

Our next task is to compute the matrix elements of $(\frac{1}{im+\hat{V}})_{nm}^{ik}$ fixing initial and final states and averaging over all coordinates of intermediate dyons. To this end we introduce as in [18] the amplitudes $D_{nm}^{ik}$ and $P_{nm}^{ik}$ for even and odd number of transitions $\hat{V}$ respectively

$$\left(\frac{1}{im + \hat{V}}\right)_{nm}^{ik} = \frac{\delta_{ik}\delta_{nm}}{im} + \begin{cases} D_{nm}^{ik}(R_i^{(i)}, R^{(k)}, \Omega^{(i)}, \Omega^{(k)}, \omega^{(i)}, \omega^{(k)}) \\ P_{nm}^{ik}(R_i^{(i)}, R^{(k)}, \Omega^{(i)}, \Omega^{(k)}, \omega^{(i)}, \omega^{(k)}) \end{cases} \tag{57}$$

In the definition (57) it is assumed that amplitudes of returns to the initial and final center $i$ ar $k$ are not included in $D^{ik}$, $P^{ik}$ and should be added separately (which makes Eq.(57) not an equality, but rather a symbolic equation).

This amplitude of the return to the center $j$ we denote as

$$\Delta_{mn} = D^{jj}_{mn}(R^{(j)}, R^{(j)}, \Omega^{(j)}, \Omega^{(j)}, \omega^{(j)}, \omega^{(j)}) \tag{58}$$

Since in $D^{jj}$ integration over all intermediate coordinates $(R^{(k)}, \Omega^{(k)}, \omega^{(k)})$ is done, $\Delta_{mn}$ does not depend on $R^{(j)}, \Omega^{(j)}, \omega^{(j)}$ and is a constant matrix.

Taking into account any number of returns to the same center $j$, brings about a matrix $\varepsilon_{mn}$, defined as:

$$\varepsilon_{mn} = \frac{1}{m}(1 - im\hat{\Delta})^{-1}_{mn} \tag{59}$$

With its help the equations, connecting $\hat{P}$ and $\hat{D}$ can be written as follows

$$P^{ik}_{nm} = -\frac{1}{im}V^{ik}_{nm}\frac{1}{im} - \frac{N}{2V_4}\int d^4R^{(j)}d\Omega^{(j)}d\omega^{(j)}\frac{1}{i}V^{ij}_{ns}\varepsilon_{sm'}D^{jk}_{m'm} \tag{60}$$

$$D^{ik}_{nm} = -\frac{N}{2V_4}\int d^4R^{(j)}d\Omega^{(j)}d\omega^{(j)}\frac{1}{i}V^{ij}_{ns'}\varepsilon_{s's}P^{jk}_{sm} \tag{61}$$

As a next step we separate out the dependence of $\hat{P}, \hat{D}, \hat{V}$ on lower indices and on $\Omega^{(i)}, \Omega^{(k)}$. To this end we consider zero-mode solutions $u^{(i)}_n(x)$ in the form of (A.11) and make Fourier transform

$$u^{(i)}_n(p) = \int u^{(i)}_n(x)e^{ipx}d^4x = e^{ip_0 2\pi n}\bar{u}^{(i)}(p) \tag{62}$$

It is important that $\bar{u}^{(i)}(p)$ does not depend on $n$ altogether. Therefore with the help of (56) one has

$$V^{ij}_{nm}(R^{(i)}, \Omega^{(i)}, \omega^{(i)}; R^{(j)}, \Omega^{(j)}, \omega^{(j)}) = \int \frac{d^4p}{(2\pi)^4}e^{ip(R^{(i)}-R^{(j)})}v^{ij}_{nm}(p), \tag{63}$$

$$v^{ij}_{nm}(p) = e^{-2\pi i(p^i_0 - p^j_0 m)}\bar{u}^+(p^i)\Omega^{+(i)}(-\hat{p})\Omega^{(j)}\bar{u}(p^j) \tag{64}$$

where $p^i = \Re_{\omega_i} p$, and $\Re_{\omega_i}$ is $0(4)$ rotation transforming time unit vector into $\omega_i$.

We introduce now "amputated" amplitudes $d, f, w$ instead of $\hat{D}, \hat{P}, \hat{V}$ as follows

$$D^{ik}_{mn} = \int \frac{d^4p}{(2\pi)^4}e^{ip(R^{(i)}-R^{(k)})-i2\pi(p^i_0 m - p^k_0 n)}\bar{u}^+(p^i)\Omega^{+(i)}d(p^i, p^k)\Omega^{(k)}\bar{u}(p^k) \tag{65}$$

and similarly for $f(p^i, p^k)$; according to (64)) one has $w(p^i, p^k) \equiv -\hat{p}$

Insertion of these definitions into Eqs.(60-61) yields

$$f(p^i, p^k) = -\frac{w(p)}{(im)^2} - \frac{N}{2V_4 N_c} \frac{w}{i} \int \nu(p^j) d\omega^{(j)} d(p^j, p^k) \tag{66}$$

$$d(p^i, p^k) = -\frac{N}{2V_4 N_c i} w(p) \nu(p^j) d\omega^{(j)} f(p^j, p^k) \tag{67}$$

where we have introduced

$$\nu(p) = \sum_{n,s} e^{+ip_0 2\pi n} \bar{u}(p) \varepsilon_{ns} e^{-ip_0 2\pi s} \bar{u}^+(p) \tag{68}$$

One can see in (67-68) that $f$ and $d$ do not depend on rotations in $p^i, p^k$ and the integration over $d\omega^i$ there acts only on $\nu(p^j)$, so that with the definition

$$\bar{\nu}(p) = \int \nu(p_j) d\omega^j \tag{69}$$

one obtains

$$d(p) = \frac{\frac{iN\hat{p}\bar{\nu}\hat{p}}{2V_4 N_c m^2}}{1 + \left(\frac{N}{2V_4 N_c}\right)^2 \hat{p}\bar{\nu}\hat{p}\bar{\nu}} \tag{70}$$

and $f(p)$ is expressed through $d$ via (66). The definition (58) can be used now to obtain the selfconsistency relation, taking into account that at $m \to 0$, $\tilde{\Delta} \sim \frac{1}{m^2}$ and therefore one has

$$\Delta_{mn} \varepsilon_{ns} = \frac{i}{m^2} \delta_{ms} \tag{71}$$

as a result of insertion of (65) and (70) into (58) multiplied with $\varepsilon_{mn}$, one has

$$n_0 = \frac{2V_4 N_c}{N} \int \frac{d^4 p}{(2\pi)^4} \frac{M^2(p)}{M^2(p) + p^2} \tag{72}$$

where we have defined the average number of zero modes per dyon $-n_0$, $n_0 \approx \frac{V_4^{1/4}}{b}$, $b$ is the internal scale parameter of dyons,

We also introduced the chiral mass $M(p)$

$$M(p) = \frac{N}{2V_4 N_c} tr(\hat{p}\bar{\nu}(p)\hat{p}) = \frac{N}{2V_4 N_c} p^2 \bar{\nu}(p) \tag{73}$$

where we used the fact that $\bar{\nu}$ is averaged over all directions and should be proportional to the unit matrix in Lorentz and color space.

Eq.(72) goes over into the corresponding consistency relation for instantons [18] when $n_0 = 1$ and matrix $\hat{\varepsilon}$ becomes a number, while $\bar{u}(p)$ is the Fourier transform of the 'tHooft's zero mode [13].

The solution $d(p)$ (70) assumes the knowledge of the matrix $\varepsilon_{ns}$, while the consistency relation (72) imposes only one condition. Therefore the strategy of solution is as follows. From (65) one finds $\Delta_{mn} \equiv D^{ii}_{mn} = \int \frac{d^4p}{(2\pi)^4} e^{-2\pi i p_0 (m-n)} \bar{u}^+(p) d(p) \bar{u}(p)$ through $d(p)$. It clearly depends only on the modulus $|m - n|$. Then inverting the matrix $\Delta_{mn}$ one finds $\varepsilon_{mn}$ from (71). Finally from (68)-(69) one finds $\bar{\nu}(p)$ and inserts it into (70), defining $d(p)$ The cycle is thus completed, and should be repeated till the convergence is achieved.

One can also study another basis of zero modes, namely that of (34). In this case dependence on $\beta$ can be also extracted, indeed

$$u^{(i)}_\beta(p) = \sum_n e^{-i\beta 2\pi n} u^{(i)}_n = \sum_n e^{2\pi n i (p_0 - \beta)} \bar{u}^{(i)}(p) =$$

$$= \sum_k \delta(\beta - p_0 - k) \bar{u}^{(i)}(p) \equiv \delta_{[\beta, p_0]} \bar{u}^{(i)}(p) \tag{74}$$

where we have introduced notation $\delta_{[\beta p_0]}$, implying that $\beta$ is in the interval [0,1] and $\delta$ – function should be moderated first, introducing finite number of centers $N_0$ in the dyon ($\sum_{n=-N_0/2}^{N_0/2}$) and considering limit $N_0 \to \infty$ at the end.

In this way one obtains the same equations (65-67) for $f, d$ if the new definitions are used, e.g.

$$D^{ik}_{\beta\beta'} = \int \frac{d^4p e^{ip(R^i - R^k)}}{(2\pi)^4} \delta_{[\beta p_0]} \bar{u}^+(p^i) \Omega^{+i} d(p^i p^+) \Omega^k \bar{u}(p^k) \delta_{[\beta' p_0]} \tag{75}$$

and where in (66)-(67) now $\bar{\nu}$ is defined as

$$\bar{\nu}(p) \to \tilde{\nu}(p) = \int d\omega \bar{u}(p) \varepsilon_{[p_0, p_0]} \bar{u}^+(p) \tag{76}$$

and

$$\varepsilon_{[p_0, p_0]} \equiv \int_0^1 d\beta \int_0^1 d\beta' \delta_{[\beta, p_0]} \varepsilon_{\beta\beta'} \delta_{[\beta', p_0]} \tag{77}$$

From (75) one deduces that $\Delta_{\beta\beta'} = \delta_{\beta\beta'}\Delta(\beta)$ and hence also $\varepsilon_{\beta\beta'}$ is diagonal due to the relation

$$\varepsilon_{\beta\beta'}\Delta_{\beta'\beta''} = \frac{i}{m^2}\delta_{\beta\beta''} \tag{78}$$

and is equal to

$$\varepsilon_{\beta\beta'} = \delta_{\beta\beta'}\varepsilon(\beta) = \delta_{\beta\beta'}\frac{i}{m}\Delta^{-1}(\beta) \tag{79}$$

with

$$\Delta(\beta) = \int \frac{d^4p}{(2\pi)^4}\delta_{[\beta p_0]}\bar{u}^+(p)d(p)\bar{u}(p) \tag{80}$$

The system (70), (72), (76), (77) and (79-80) is now complete.

We now proceed to write down the quark propagator (54) in terms of functions $d, f$ and finally in terms of the chiral mass $M(p)$ (73).

Following the same procedure as in [18], one can rewrite (54) as

$$S(p) = \frac{\hat{p}}{p^2} - \frac{N}{2V_4}\left(\frac{\delta_{ns}}{im} + \left(\Delta\frac{1}{1-im\Delta}\right)_{ns}\right) \times$$

$$\times \int d\Omega^{(i)}d\omega^{(i)}(u_n^{(i)}(p,\omega^{(i)})u_s^{(i)+}(p,\omega^{(i)}) + \text{ dyon } \leftrightarrow \text{ antidyon}) -$$

$$-\left(\frac{N}{2V_4}\right)^2 \int d\Omega^{(i)}d\Omega^{(j)}d\omega^{(i)}d\omega^{(j)}u_n^{(i)}(p,\omega^{(i)})(m\varepsilon_{ns})D_{sl}^{ij}(m\varepsilon_{lk})u_k^{+j}(p,\omega^j)$$

$$-\left(\frac{N}{2V_4}\right)^2 \int d\Omega^{(i)}d\Omega^{(j)}d\omega^{(i)}d\omega^{(j)}u_n^{(\bar{i})}(p,\omega^{(i)})(m\varepsilon_{ns})D_{sl}^{\bar{i}\bar{j}}(m\varepsilon_{lk})u_k^{+\bar{j}}(p,\omega^j)$$

$$-\left(\frac{N}{2V_4}\right)^2 \int d\Omega^{(i)}d\Omega^{(j)}d\omega^{(i)}d\omega^{(j)}[u_n^{(i)}(p,\omega^{(i)})(m)\varepsilon_{ns}P_{sl}^{i\bar{j}}(m)\varepsilon_{lk}u_k^{+\bar{j}}(p,\omega^j)$$

$$+ (i \to \bar{i}, \bar{j} \to j)] \tag{81}$$

In the appendix we write explicitly the structure of the dyon zero–modes density matrix $u_n^{(i)}u_s^{(i)+}$. Submitting in (81) expression (65-67) and (70), (72) we finally obtain $S(p)$ in the form

$$S(p) = \frac{\hat{p} + iM(p)}{p^2 + M^2} \tag{82}$$

This form justifies the meaning of $M(p)$ as a chiral mass, i.e. an effective mass of quark due to CSB. It coincides with the form of $S(p)$ for the instanton gas [18], however the explicit expression for $M(p)$ (73) differs.

The most remarkable feature of (82) is the disappearance of the massless pole $\frac{\hat{p}}{p^2}$ from $S_0(p)$. One should have in mind of course that the form (82) is gauge–noninvariant and obtained neglecting confinement. If one takes into account these effects, as in [21], the pole structure in (82) is supplemented by the area law due to the string between the given quark and an antiquark and the pole is never present in physical amplitudes.

From (82) one can easily compute the chiral condensate:

$$< \bar{q}q >_{Mink.} = -i < \bar{q}q >_{Eucl.} = i < tr S(x,x) >=$$

$$= i \int \frac{d^4p}{(2\pi)^4} S(p) = -4N_c \int \frac{d^4p}{(2\pi)^4} \frac{M(p)}{p^2 + M^2(p)} \quad (83)$$

It is nonzero thus confirming the phenomenon of CSB in the dilute dyonic gas.

# 7  Conclusions and discussions

We have described two different infinite sets of quark zero modes for an (anti) dyon. We suggest to consider these zero modes as a driving mechanism for chiral symmetry breaking (CSB) in a balanced dyon–antidyon ($d\bar{d}$) gas, similarly to the CSB mechanism in the instanton gas [18]. The composition of the $d\bar{d}$ gas is by itself a nontrivial problem because of the long-distance field of dyons. We have found one specific gauge where the superposition ansatz of the $d\bar{d}$ vector potentials provides a finite action.

The resulting $d\bar{d}$ gas has classical Coulombic – type interaction, so that dyons and antidyons are not confined. Then according to the 't Hooft's principle [26] one may expect Confinement of quarks (and gluons). The possible role of dyons in the Confinement mechanism was also emphasized in Ref.[28]. The numerical studies of the Wilson loop in the $d\bar{d}$ gas are now in progress [25]. But the main emphasis of the present paper is on the possibility of CSB in the $d\bar{d}$ gas. Indeed, if the CSB is present in the $d\bar{d}$ gas then one may hope to have a model of the QCD vacuum, which obeys both properties of confinement and CSB. At the same time, it gives a possibility of explanation of the simultaneous disappearance of confinement and chiral symmetry above the critical temperature. A similar proposal made by one of us [27] attributes Confinement to a different kind of multiinstanton configurations. Our result might well be extendible to this model as well, since the main ingredients of our formulas are quite general. In section 6 we have calculated the one-quark propagator in the balanced $d\bar{d}$ gas and demonstrated the appearance of the chiral quark mass.

The selfconsistency equation (72) has the form similar to the case of the instanton gas [18], and goes over into the latter when the number of zero modes on each center (dyon or instanton) tends to unity. Also the form of the quark propagator is similar and it is clear that the quark–antiquark Green's functions can be obtained by the same method, as used for the instanton gas in [18].

From (82)-(83) (and the corresponding $q\bar{q}$ propagators) one can see that for $N_f > 1$ the original symmetry $U_L(N_f) \otimes U_R(N_f)$ of the QCD Lagrangian is broken down to $SU_V(N_f) \otimes U_V(1)$, while both $SU_A(N_f)$ and $U_A(1)$ are broken. Thus the topological charges of dyons, creating fermion zero modes, produce a universal mechanism for breaking both $U_A(1)$ and $SU_A(N_f)$, found for instantons by 'tHooft [15] and Diakonov and Petrov [18]. We refer to this universal mechanism as the topological mechanism of CSB.

Many of features of the problem under study are left unclarified.

First, and the most important is the detailed numerical study of the $d\bar{d}$ gas with classical interaction and quantum corrections to yield the equilibrium density of gas and resulting numbers for string tension and chiral condensate.

This question is now under study.

Second, the effective action for large $N_f$ and the bosonization procedure is not done in the paper. Therefore, the effective action for Goldstone–Nambu bosons is lacking above. This is planned for future publications.

The most part of this work was done while the second author (Yu.S.) has been visiting the theoretical department of the Universidad Autónoma de Madrid. This author is grateful for kind hospitality to the members of theoretical department and all the staff. He also wants to thank P. van Baal and G. 't Hooft for useful dicussions.

He was also partly supported by the Russian Fund for Fundamental Research, Grant 93-02-14937. A. G-A acknowledges finantial support by the CICYT grants AEN93-0693 and the EC network CHRX-CT93-0132.

**Appendix**

In the appendix we give a short derivation of the solution of the Dirac equation in the background field of a multi-instanton solution of the 't Hooft ansatz form (Eq. (8)). Then we will specialize to the case of the dyon solution.

Consider the field tensor $F_{\mu\nu}$ constructed from a gauge potential $A_\mu$ of 't Hooft 's form. One can separate it into a self-dual and anti-self-dual parts $\omega_i^{(+)}$ and $\omega_i^{(-)}$ as shown in Eq. (18). In terms of these parts one can compute the product of the modified Dirac operators $\mathcal{D}$ as follows:

$$\hat{\mathcal{D}}\hat{\bar{\mathcal{D}}} = \mathcal{D}_\mu \mathcal{D}_\mu - 2i\omega_i^+ \Gamma_{is} \tag{A.1}$$

$$\hat{\bar{\mathcal{D}}}\hat{\mathcal{D}} = \mathcal{D}_\mu \mathcal{D}_\mu - 2i\omega_i^- \bar{\Gamma}_{is} \tag{A.2}$$

For purely self-dual fields $\omega_i^{(-)} = 0$ and one sees that there are no negative chirallity zero-modes. Thus, we may simply consider the equation

$$\hat{\bar{\mathcal{D}}}\chi^{(+)}(x) \equiv (\hat{\bar{\partial}}_s - 2P_+(\hat{\partial}_c f))\chi^{(+)}(x) = 0 \tag{A.3}$$

Now let us write the following ansatz for $\chi^{(+)}(x)$

$$\chi^{(+)}(x) = (\hat{\partial}_s \phi)u_+. \tag{A.4}$$

The ansatz follows from acting with $\hat{\partial}_s$ on Eq.A.3 and subsequently with the inverse of the 4-dim Laplace operator $\Delta^{-1}$. Substituting the form A.4 into A.3., we get

$$(\Delta\phi)u_+ = 2P_+((\hat{\partial}_c f)(\hat{\partial}_s \phi)u_+ \tag{A.5}$$

Now using the fact that

$$P_+ \Gamma_{\alpha c} \Gamma_{\beta s} u_+ = \delta_{\alpha\beta} u_+ \tag{A.6}$$

which follows from Eq. (17), we obtain the following equation for $\phi$:

$$\Delta\phi = 2\partial_\mu f \partial_\mu \phi. \tag{A.7}$$

Now we can rewrite the previous as follows:

$$e^f \Delta(\phi e^{-f}) = \phi(e^f \Delta e^{-f}) \tag{A.8}$$

The left hand side of the previous equation vanishes as a consequence of the self-duallity of the gauge field configuration. Hence, the Laplace operator acting on the product of $\phi$ and $e^{-f}$ vanishes except whenever $e^f$ vanishes. Taking for $F \equiv e^{-f}$ the form (22), we conclude that $\phi$ must be a linear combination of the following functions

$$\phi^{(k)}(x) = \frac{e^f}{(x - x_{(k)})^2} \tag{A.9}$$

Therefore there are $N$ linearly independent zero modes, in agreement with the index theorem.

If we use the form Eq. (23), then again Eq. (A.9) gives solutions of the Dirac equation, but this time only $N-1$ are linearly independent, as it should.

In general given $W \equiv F^{-1} = e^f$ and letting $x_{(k)}$ (for $k = 1, ..., N$) be the position of its zeros, we can write the most general solution of the Dirac equation ($\hat{\tilde{D}}\psi^{(+)} = 0$) as follows:

$$\psi_d^{(+)} = \sum_{k=1}^{N} d_{(k)} F^{1/2} \hat{\partial} \Big( \frac{F^{-1}}{(x - x^{(k)})^2} \Big) u_+ \tag{A.10}$$

where $d_{(k)}$ are arbitrary complex constants. In obtaining Eq. (A.10) we have combined Eq. (11), Eq. (A.4) and Eq. (A.9). These solutions coincide with those found by Grossman [24].

Concerning the normalization, one must again distinguish the cases given by formulas (22) and (23). In the first case, after some algebra one can show that choosing $d_{(k)} = \frac{\rho_k}{2\pi^2} \delta_{ki}$ is an orthonormal basis of the space of zero-modes for $i = 1, \ldots N$. In the second case, one must form an unitary transformation of this basis, orthogonal to the vector $d_{(k)} = \rho_k^2$ which is trivial. Hence, there are only N-1 elements of the basis.

Now let us specialize to the dyon solution Eq.(24). The total topological charge diverges and there are infinitely many zeros of $F_{HS}$. Hence, there is also an infinite number of zero-modes. A possible non-orthogonal basis of the space of solutions is then:

$$u_k(x) = F^{1/2} \hat{\partial} \Big( \frac{F^{-1}}{(x - x^{(k)})^2} \Big) u_+ \tag{A.11}$$

Since both $F$ and $W$ are periodic in time with period $2\pi$ it is natural to consider the different independent zero modes to be independent Fourier components:

$$\psi^{(\beta)}(x_0 + 2\pi) = e^{i\beta 2\pi}\psi^{(\beta)}(x_0) \qquad (A.12)$$

Formula (33) and (35) of Section 3 summarizes these results. Notice that $\beta$ ranges from 0 to 1, but the value of $\beta = 0$ gives a trivial solution, a consequence of the same phenomenon which occurs for $e^{-f}$ of the form (23). All solutions are orthogonal to each other.

To conclude let us write down the expression for the projector onto the zero-modes. Using $\alpha$ and $\alpha'$ for the 4-spinor indices and $a, a'$ for the colour ones we may write

$$\psi^{(\beta)}_{\alpha a}\psi^{(\beta')+}_{\alpha' a'} = \frac{F}{2\pi^2}\partial_\mu(\frac{F^{(\beta)}}{F})\partial_\nu(\frac{F^{(\beta')}}{F}))\ (\bar{\Gamma}_\mu\bar{\Gamma}_\tau\Gamma_\nu)_{aa'}(\gamma_0\gamma_\tau\frac{I+\gamma_5}{2})_{\alpha\alpha'} \qquad (A.13)$$

The equation follows from the expression of the zero-mode and by using Eq (12) for $P_+$.